\documentclass[twocolumn,pra,aps,superscriptaddress,showpacs,amsmath,amssymb,floatfix]{revtex4-1}
\usepackage{graphicx}
\usepackage{amssymb}
\usepackage{amsmath}
\usepackage{epsfig}
\usepackage{color}
\usepackage{mathtools}
\usepackage[colorlinks,linkcolor=blue,anchorcolor=blue,citecolor=blue,urlcolor=blue]{hyperref}
\usepackage{physics}
\usepackage{ulem}

\begin{document}

\title{Hermitian and Non-Hermitian Topological Transitions Characterized by Manifold Distance}

\author{ZhaoXiang Fang}
\email{ZhaoXiang Fang and Ming Gong contribute equally}
\affiliation{School of Physical Science and Technology, Xinjiang University, Urumqi, 830046, China}

\author{Ming Gong}
\affiliation{Key Lab of Quantum Information, Chinese Academy of Sciences, School of physics, University of Science and Technology of China, Hefei, 230026, China}
\affiliation{Synergetic Innovation Center of Quantum Information and Quantum Physics, University of Science and Technology of China, Hefei, 230026, China}

\author{Guang-Can Guo}
\affiliation{Key Lab of Quantum Information, Chinese Academy of Sciences, School of physics, University of Science and Technology of China, Hefei, 230026, P.R. China}
\affiliation{Synergetic Innovation Center of Quantum Information and Quantum Physics, University of Science and Technology of China, Hefei, 230026, P.R. China}

\author{Yongxu Fu}
\email{yongxufu@pku.edu.cn}
\affiliation{International Center for Quantum Materials, School of Physics, Peking University, Beijing, 100871, China}

\author{Long Xiong}
\email{XiongLong_phy@outlook.com}
\affiliation{International Center for Quantum Materials, School of Physics, Peking University, Beijing, 100871, China}

\date{\today }

\begin{abstract}
Topological phases are generally characterized by topological invariants denoted by integer numbers. However, different topological systems often require diffenent topological invariants to measure, and theses definition usually fail at critical points. Therefore, it's challenging to predict what would occur during the transformation between two different topological phases. To address these issues, we propose a general definition based on fidelity and trace distance from quantum information theory: manifold distance (MD). This definition does not rely on the berry connection but rather on the information of the two manifolds - their ground state wave functions. Thus, it can measure different topological systems (including traditional band topology models, non-Hermitian systems, and gapless systems, etc.) and exhibit some universal laws during the transformation between two topological phases. Our research demonstrates: for different topological manifolds, the change rate (first-order derivative) or susceptibility (second-order derivative) of MD exhibit various divergent behaviors near the critical points. Compared to the strange correlator, which could be used as a diagnosis for short-range entangled states in 1D and 2D, MD is more universal and could be applied to non-Hermitian systems and long-range entangled states. For subsequent studies, we expect the method to be generalized te real-space or non-lattice models, in order to facilitate the study of a wider range of physical platforms such as open systems and many-body localization.
\end{abstract}

\maketitle

In the literature of topology and geometry, two manifolds are topologically equivalent when and only when they can be smoothly deformed to each other \cite{lee2010introduction}. The equivalent manifolds are uniformly characterized by one topological index (invariant), such as the Chern number \cite{lee2010introduction} corresponding to the famous Thouless-Kohmoto-Nightingale-den Nijs (TKNN) number derived from the linear response theory in the theory of topological matters \cite{hatsugai1993chern,kane2005z}. This pioneering breakthrough propelled the establishment of the topological classification of matters with various symmetries in all dimensions \cite{qi2011topological,hasan2011three,qi2011topological,schnyder2009classification,schnyder2008classification}. Regardless of the Hermiticity of the topological matters, theoretical explorations of non-Hermitian quantum systems have significantly expanded the scope of condensed matter physics in the past decade \cite{ashida2020non,bergholtz2021exceptional,gong2018topological,kawabata2019symmetry,shen2018topological,yao2018edge,song2019non,lieu2018topological,bagarello2016non,PhysRevLett.116.133903,kunst2018,yao201802,lee2019an,yokomizo2019,origin2020,slager2020,yang2020,zhang2020,xue2021simple,guo2021exact,edgeburst2022,fu2023ana}, rapidly encompassing higher-order non-Hermitian systems \cite{kawabata2019second,lee2019ho,edvardsson2019,kawabatahigher,okugawa2020,fu2021,yu2021ho,palacios2021,st2022}, exceptional points \cite{kawabata2019,yokomizo2020,jones2020,zhang2020ep,xue2020dirac,yang2021,denner2021,fu2022,mandal2021ep,delplace2021ep,liu2021ep,marcus2021ep,ghorashi2021dirac,ghorashi2021weyl}, and scale-free localization \cite{li2021impurity,libo2023scale,guo2023scale,fu2023hybrid,molignini2023anomalous}. In open boundary conditions (OBSs), the non-Hermitian skin effect (NHSE) is a remarkable feature that predicts an extensive number of eigenstates localized at the edges as well as the breakdown of the Bloch band theory \cite{PhysRevLett.116.133903, yao2018edge,song2019non,yokomizo2019,zhang2020,yang2020,origin2020}. A comprehensive consequence is the difference of the topological transition points between OBCs and periodic boundary conditions (PBCs). 

However, a subtle question has never been addressed in either Hermitian or non-Hermitian systems: what quantitative contexts will happen during the deformation of the transition of two manifolds (topological phases)? Note that in quantum information science, there are two common ways to measure the similarity between two pieces of information: trace distance and fidelity (in the case of pure states, these are completely equivalent) \cite{jozsa1994fidelity,nandi2018two,liang2019quantum,gu2010fidelity,banchi2015quantum,li2012superfidelity,rastegin2007trace,rastegin2007trace,zhang2019subsystem,de2023subsystem,liang2019quantum,brito2018quantifying,rana2016trace}. Based on this concept, various new definitions emerge for different physical systems, such as the fidelity rate to characterize the quantum phase transition of the ground state \cite{gu2010fidelity,banchi2015quantum}, the trace distance quantum discord to measure the quantum correlation \cite{rana2016trace}, and the minimum trace distance to quantify the non-locality of Bell-type inequalities \cite{brito2018quantifying}. These generalized concepts are built upon the significant distinctions in “distances” between different phases of matter. Thus, the measurement of the quantum state serves as an inspiration for our investigation of the topological phase boundaries \cite{wiseman2009quantum,zeng2019quantum}.

In this work, we establish a formulation to analyze the divergent behavior observed during the deformation of two manifolds representing topological phases. We introduce the concept of “manifold distance (MD)” as a tool for efficiently and directly identifying the boundaries of topological phases. We find that MD, along with its higher-order derivatives, transitions smoothly between two topologically equivalent manifolds (phases). However, at critical points separating distinct topological phases, the higher-order derivatives of MD exhibit distinct divergent behaviors governed by universal scaling laws. 

\begin{figure}
	\centering
	\includegraphics[width=0.48\textwidth]{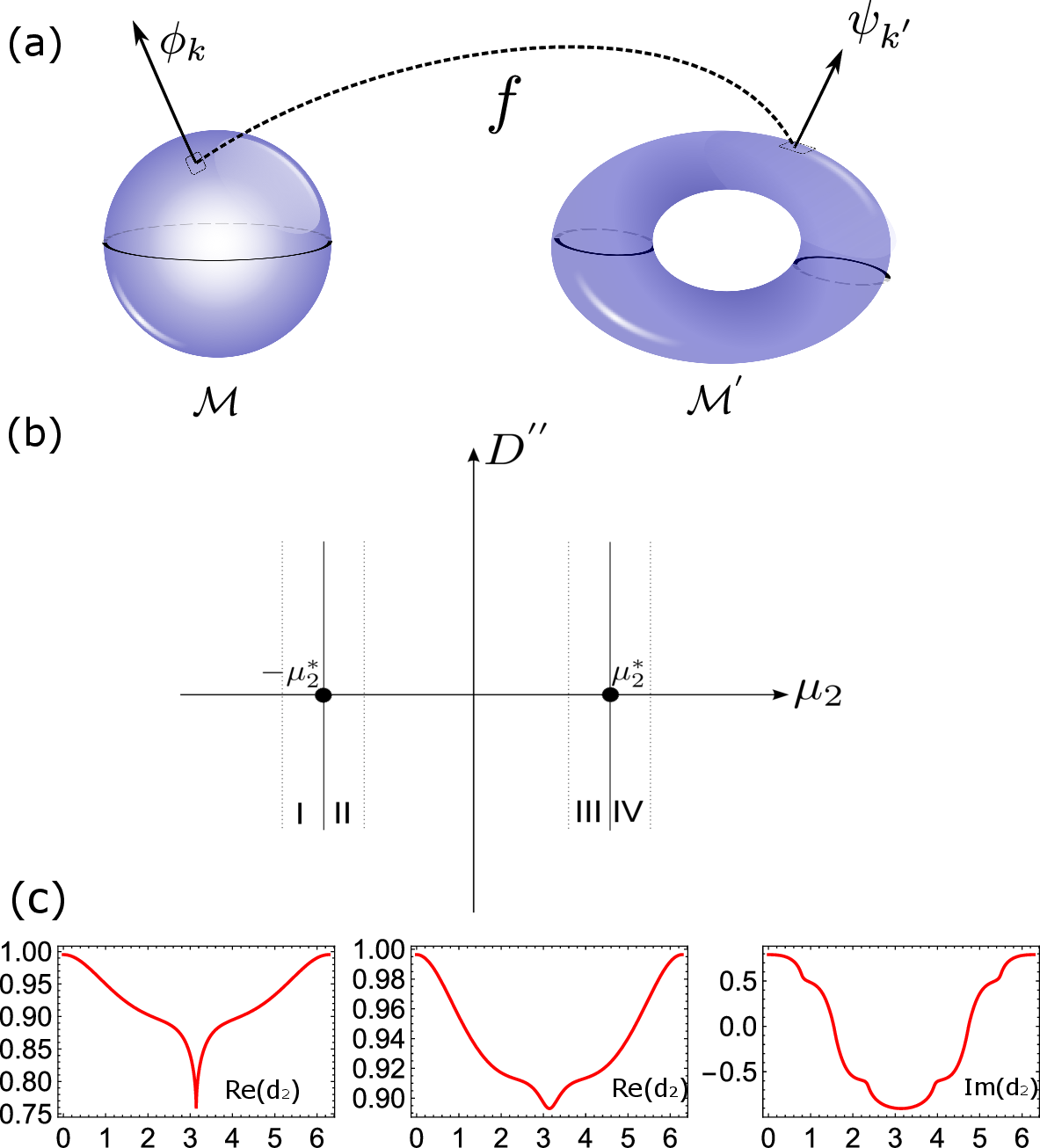}
	\caption{Adiabatic deformation between two manifolds. (a) In this work, the parameters of the two base spaces $\mathcal{M}$ and $\mathcal{M}^{\prime}$, denoted as $k$ and $k^{\prime}$, are connected by a smooth function $f: k \in \mathcal{M} \rightarrow k^{\prime} \in \mathcal{M}^{\prime}$. (b) For most non-Hermitian models discussed, the phase regions, labeled I, II, III, and IV (from left to right), are separated by black dots representing critical phase transition points. (c) The integrand of the manifold distance $d_2$ in Eq. (2) shows a singularity in its real part at phase transition points within the Brillouin zone, while remaining smooth elsewhere. This singularity leads to divergence in the derivative of the manifold distance at phase boundaries. The imaginary part of $d_2$ is either zero or an even function over the integration range, making no contribution to the manifold distance.}
	\label{fig1}
\end{figure}

Our MD formulation is broadly applicable, extending not only to conventional Hermitian systems but also to non-Hermitian systems under both open boundary conditions (OBCs) and periodic boundary conditions (PBCs). Furthermore, we demonstrate the applicability of our method to (many-body) continuous systems, including $p$-wave superconductors and the Kitaev toric code model, which exhibits topological order. This approach provides a rigorous means to characterize and understand the behavior at topological phase boundaries. A pictorial representation of this process is shown in Fig. \ref{fig1}.

To illustrate the formulation, consider two manifolds, $\mathcal{M}$ and $\mathcal{M}'$, whose parameters are mapped onto each other via a smooth function: 
\begin{equation} 
f: {\bf k} \rightarrow {\bf k}' = f({\bf k}). 
\end{equation} 
The mapping functions, which quantify the distance between different ground-state wavefunctions, exhibit divergent behavior at the critical points of phase transitions.

%Because of the two different uses, the connection based on wave functions at the same point is no longer valid. One needs to seek a more general definition of distance, instead of connection, to measure this deformation process. 

%In physical models when the parameter ${\bf k}$ may refer to the momentum, which is always a good quantum number[]. 

%This momentum is assumed to be unchanged during deformation. However, in this work, we consider all possible mappings. We will show that for 

{\it Manifold distance.-} The most natural definition of manifold distance is based on the trace distance\cite{jozsa1994fidelity}, which, for pure states, is exactly equivalent to fidelity \cite{jozsa1994fidelity,liang2019quantum}. We define two distinct distance measures as follows:

\begin{equation} 
d_1 = 1 - |\langle \psi_k|\phi_{k'}\rangle|^2, \quad d_2 = \sqrt{1 - |\langle \psi_k|\phi_{k'}\rangle|^2}. \label{d1d2} 
\end{equation}

It is evident that $d_i \geq 0$ and $d_1 \leq d_2$.

Our primary focus is on pure states, although the same definitions could be extended to mixed states. In such cases, the overlap between two distinct regions is determined by the distance between the corresponding density matrices \cite{jozsa1994fidelity}. To quantify the distance between two manifolds, we define the manifold distances as:
\begin{equation} 
	\mathcal{D}_1 = \int d{\bf k} , d_1({\bf k}, {\bf k}'), \quad
	\mathcal{D}_2 = \int d{\bf k} , d_2({\bf k}, {\bf k}'). 
\end{equation}
A distance $\mathcal{D}_i = 0$ implies that, for every ${\bf k}$ and ${\bf k'}$, the two wavefunctions are identical.

The concept of manifold distance can be intuitively understood as follows: Consider the transformation $|{\bf k}'| = |{\bf k}| + c$, representing a mapping within the same manifold, as illustrated in Fig. \ref{fig1}(d). For $|{\bf c}| \in \mathbb{R}$, we find that $\langle \psi_k|\psi_{k+c}\rangle = 1 + c \langle \psi_k|\partial_k|\psi_k\rangle = 1 + ic A_k$, where $A_k = -i\langle \psi_k|\partial_k|\psi_k\rangle$ defines the Berry connection. In this case, the distances are given by:
\begin{equation}
	d_1 = 1 - |1 + ic A_k|^2 = c^2 A_k^2, \quad d_2 = |c A_k|.
\end{equation}
If $c$ is purely imaginary, a similar calculation yields $d_1 = ic A_k$ and $d_2 = \sqrt{|c A_k|}$. 

During a topological phase transition, the energy bands of the system can undergo significant changes at certain points in the Brillouin zone, resulting in a divergence in the manifold distance. To illustrate these properties, we examined a simplified non-Hermitian Hamiltonian. A detailed theoretical analysis is provided in \textcolor{red}{Appendix VII} \cite{Supplemental}.  

In comparison to the strange correlator—an established diagnostic tool for short-range entangled states in 1D and 2D systems \cite{shankar2011equality,you2014wave,wierschem2014strange,lepori2023strange,you2014topological}—manifold distance is more universal. It applied to non-Hermitian systems, long-range entangled states (as demonstrated in Fig. 4), and even higher-dimensional or higher-order topological systems \textcolor{red}{(results forthcoming)}.

\begin{figure}
	\centering
	\includegraphics[width=0.48\textwidth]{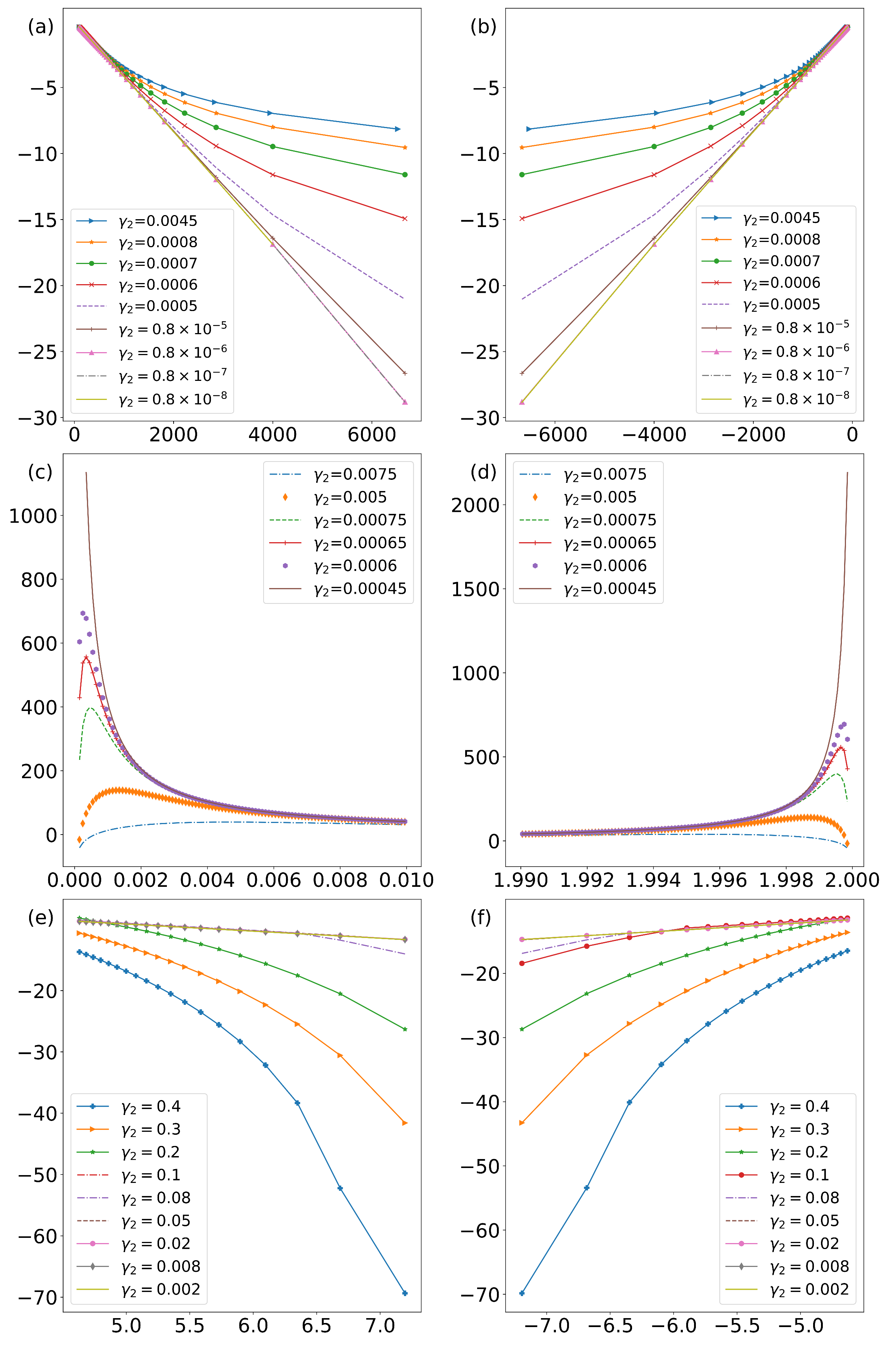}
	\caption{Non-Hermitian systems of manifold distance $D$ and its second derivative $D^{\prime \prime}$.
Fig (a) to (d) show a one-dimensional topological model from Eq. (7), illustrating the divergence of $D^{\prime \prime}$ in (a) and (b). As the non-Hermitian term coefficient $\gamma_2$ decreases, the divergence gradually returns to the Hermitian case. Fig (a) and (d) correspond to regions (I) and (IV) of the phase diagram in Fig. 1, respectively, (e) and (f) show the divergence of $D^{\prime \prime}$ for a two-dimensional topological model, corresponding to regions (I) and (II) of the phase diagram in Fig. 1.}
	\label{fig2}
\end{figure}

{\it Generalized to the Non-Hermitian.-}Non-Hermiticity is typically introduced by incorporating non-Hermitian (NH) hopping terms and/or NH gain/loss terms \cite{moiseyev2011non,liu2020gain,ashida2020non,bagarello2016non,kawabata2019symmetry}. Furthermore, certain topological invariants of Hermitian Hamiltonians are generalized when the Hermiticity condition is relaxed \cite{bergholtz2021exceptional,gong2018topological,shen2018topological,yao2018edge,song2019non,lieu2018topological}. Consequently, extending the concept of manifold distance to non-Hermitian systems is a natural progression.

Consider the eigenvalue equations of Non-Hermitian Hamiltonian $H_k$:
\begin{align}
	H_k \ket{\varphi_n} = E_n \ket{\varphi_n},\quad H_k^{\dagger} \ket{\phi_n} = E_{n}^{'*} \ket{\phi_n}.
\end{align}
Here we have four choices
\begin{equation}
    d_1 = 1 - | _1\langle a_k| b_{k'}\rangle_2 |^2, \quad 
	d_2 = \sqrt{1 - | _1\langle a_k| b_{k'}\rangle_2 |^2}. 
	\label{d_eq}
\end{equation}
where $a_k$ and $b_k$ correspond to $\psi_k$ and $\phi_k$, respectively. 

Thus, irrespective of the specific definition of manifold distance, the behavior of the phase boundary is effectively captured, with differences manifesting only in the numerical values.

In the context of non-Hermitian systems, it is advisabled to employ forms such as $\langle \psi_k|\psi_{k'}\rangle$ or $\langle \phi_k|\phi_{k'}\rangle$ to address normalization issues.

{\it Hamiltonians.-} We investigated the divergent properties of phase boundaries using one-dimensional and two-dimensional topological models \cite{kitaev2001unpaired,greiter20141d,leijnse2012introduction,gunter2005p,leijnse2012introduction,greiter20141d,chhajed2020ising,thakurathi2014majorana,Sato2016,ren2016topological}. The Hamiltonians considered are as follows:
\begin{equation}
	H_1(k) = \begin{pmatrix}
		\epsilon_k & \beta_k \\
		\beta_k  & - \epsilon_{k}
	\end{pmatrix}, \quad H_2({k}) = \begin{pmatrix}
		\epsilon_{k} & \beta_{k} \\
		\beta_{k}^*  & - \epsilon_{{k}}
	\end{pmatrix}.
	\label{use_h}
\end{equation}

In case of $H_1$, we define $\epsilon_k=-2 t \cos (k)+i \gamma$ and $\beta_k=\alpha \sin (k)$. For $H_2$, we set $\epsilon_{\mathbf{k}}=-2 t\left(\cos \left(k_x\right)+\cos \left(k_y\right)\right)-\mu+i \gamma$ and $\beta_{\mathrm{k}}=\alpha\left(\sin \left(k_x\right)+i \sin \left(k_y\right)\right)$, where $\mu$ represents the chemical potential, $t$ denotes the hopping amplitude between neighboring sites, $\alpha$ could be interpreted as the spin-orbit coupling or superconducting pairing strength, and $\gamma$ represents the non-Hermitian term coefficient.

These two models are representative of various topological phases. For instance, $H_1$ can be interpreted as the Kitaev toy model. Similarly, $H_2$ could be associated with a two-dimensional superconducting model.

%The topological phase transitions in these two models and the divergent behaviors of geometry phases have been well-known. 

Let us first analyzed the Hamiltonian $H_1$. We assume that the parameters for the manifold $\mathcal{M}$ are $\mu, t$, and $\alpha$, while for the corresponding manifold $\mathcal{M}^{\prime}$, the parameters are $\mu^{\prime}, t^{\prime}$, and $\alpha^{\prime}$. Based on the expression for $d_a$, we found that singular points occur at $k=0$ when $\mu=2 t \sqrt{1-\gamma^2 / \alpha^2}$, or at $k=\pi$ when $\mu=-2 t \sqrt{1-\gamma^2/ \alpha^2}$.
Similarly, for the Hamiltonian $H_2$, the singular points are located at $\mu= \pm 2 t\left(1+\sqrt{1-\gamma^2 / \alpha^2}\right)$.

{\it Transition from non-Hermitian to Hermitian.-} As for the above Non-Hermitian Hamiltonian $H_1$ and $H_2$, when $\gamma \to 0$, they both reback to Hermitian systems. This transition phenomenon can be demonstrated using the manifold distance. For instance, the divergence behavior of $D^{\prime \prime}$ at the phase boundary would reverts from non-Hermitian to Hermitian systems, as shown in Fig. (\ref{fig2}).

For the Non-Hermitian Hamiltonians $H_1$ and $H_2$ discussed above, as $\gamma \to 0$, both systems reduce to their Hermitian systems. This transition could be effectively characterized using the manifold distance. Specifically, the divergence behavior of $D^{\prime \prime}$ at the phase boundary transitions from the non-Hermitian regime to the Hermitian regime, as illustrated in Fig. (\ref{fig2}).

%First, although $D$ can be smooth and continuous through the entire Brillouin zone, $D^{\prime}$ passes continuously but not smoothly through the critical point, and $D^{\prime \prime}$ appears to diverge near the phase boundary. These regulations are consistent across different definitions as eq. (\ref{d_eq}) and (\ref{d_eq2}), also independent of the base point of momentum $k$ in fig. 2.

Consider a simplified 1D Hamiltonian:
\begin{align}
	H_k=\left(\begin{array}{cc}
		-\mu+i \gamma & \alpha k \\
		\alpha k & \mu-i \gamma \\
	\end{array}\right),
    \label{Hk-D}
\end{align}
\begin{figure}
	\centering
	\includegraphics[width=0.48\textwidth]{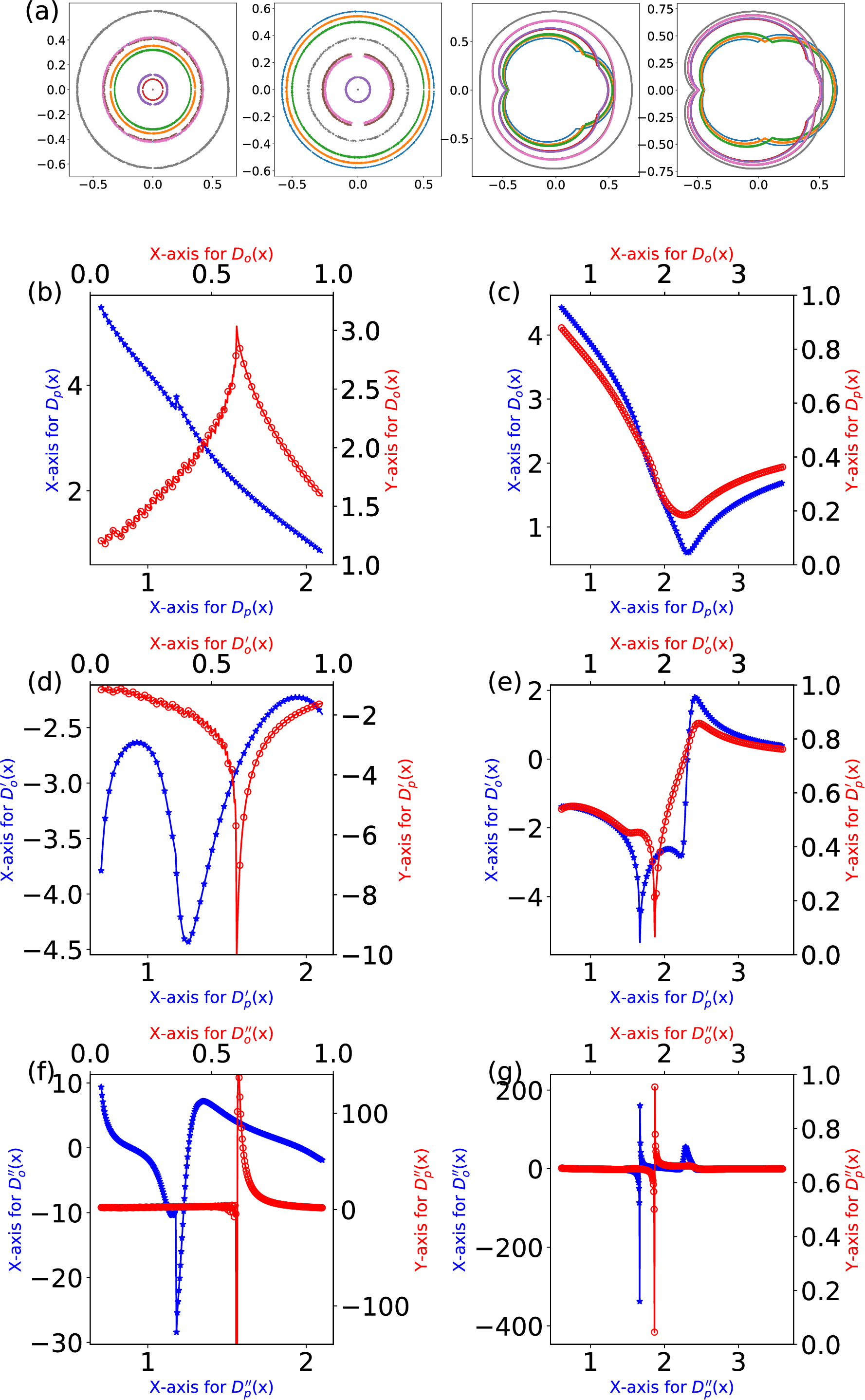}
	\caption{Manifold distance for systems with OBC. Fig (a) shows the generalized Brillouin zone (GBZ). The first two subplots represent systems without next-nearest-neighbor interactions, where the GBZ is circular, while the latter two include next-nearest-neighbor interactions, forming a GBZ with two curved segments. Parameters are chosen near the phase transition point: $t=\sqrt{\left(t^{\prime}\right)^2+\left(\frac{\gamma}{2}\right)^2}$, $t=t^{\prime}+\frac{\gamma}{2}$. The variation in the GBZ image does not capture the phase transition process. Fig (b) to (g) compare the OBC (left column) and PBC (right column) cases. Red and blue curves correspond to systems with and without next-nearest-neighbor interactions, respectively. For $t_3 = 0$, the manifold distance derivative diverges at phase boundaries in Eq. (15) (PBC) and Eq. (17) (OBC), with phase transitions at $t \approx 1.20$ and $t \approx 1.67$, consistent with Ref. \cite{yao2018edge}. For $t_3 \neq 0$, while we lack an analytical expression for the phase boundary, we can numerically identify it from the singularity in manifold distance. For OBCs, we find the phase transition at $t \approx 1.56$, in agreement with Ref. \cite{yao2018edge}, as shown in the numerical spectra of the real-space Hamiltonian for an open chain.}
	\label{fig3}
\end{figure}
\begin{figure}
	\centering
	\includegraphics[width=0.47\textwidth]{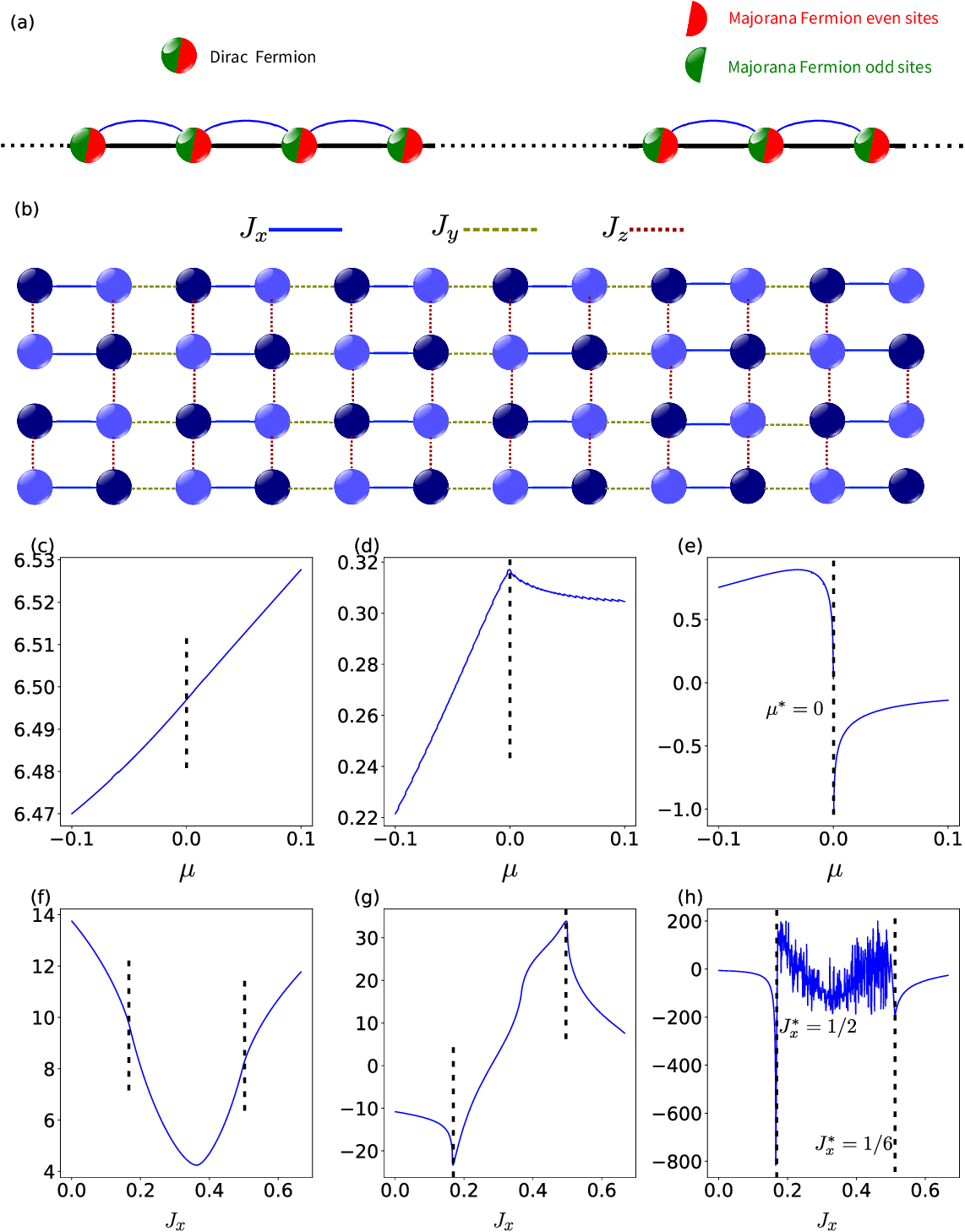}
	\caption{(a) Schematic of the Kitaev p-wave superconducting model, where Majorana fermions emerge at the ends of the chain under certain conditions, driving the system into a topological superconducting phase. (b) The Kitaev toric model, where the honeycomb lattice can be deformed into a brick-wall lattice without changing its topology. Fig (c) to (e) show the manifold distance for the p-wave SC model, with a phase transition occurring at $\mu = 0$, where $\mu$ is the chemical potential.
Fig (f) to (h) display the manifold distance for the Kitaev toric model, which has two distinct phases in its ground state: a gapped phase with Abelian anyon excitations, and a gapless phase with non-Abelian anyons. For the condition $J_x + J_y + J_z = 1$, the system exhibits these two phases. By choosing $J_x$ as the independent variable and setting $J_z = \frac{1}{3}$, two phase transition points are identified at $J_x = \frac{1}{6}$ and $J_x = \frac{1}{2}$, consistent with the derivative of $D$. Further details are available in the supplementary materials \cite{Supplemental}.}
	\label{fig4}
\end{figure}
Numerical calculations revealed that the divergence coefficients are primarily influenced by a specific set of parameters. For simplicity, we assumed the first set of parameters to be constants and omit the subscripts in subsequent discussions. Since the simplified Hamiltonian lacks a Brillouin zone, we only need to consider integration intervals that encompass the singularities of $D^{\prime \prime}$, where $D^{\prime \prime}$ denotes the partial derivative of $D$ with respect to the chemical potential. When the non-Hermitian term $i \gamma$ is large $(\gamma \gg \mu)$, the divergence behavior of $D^{\prime \prime}$ is approximately given by:
\begin{align}
	D^{''} \approx -\frac{\sqrt{\gamma}}{2 \alpha \gamma} \frac{1}{\sqrt{\mu}}, \gamma \gg \mu.
\end{align}
As for $\gamma \to 0$, the divergence behavior transitions to:
\begin{align}
    D^{''} \propto \frac{1}{\mu \sqrt{\alpha^2+\mu^2}} \approx \frac{1}{\mu \alpha},
\end{align}
which corresponds to the behavior of a Hermitian system.
This result shows that as the non-Hermitian term $i \gamma$ gradually vanishes, the divergence behavior transitions from $\frac{1}{\sqrt{\mu}}$ to $\frac{1}{\mu}$, with a superposition of divergence behaviors observed during this process. Coefficients for additional models could be found in Section IX of the \textcolor{red}{supplementary materials} \cite{Supplemental}.

Finally, although the divergence behavior of $D^{\prime \prime}$($D^{\prime}$) is affected by several parameters, its divergence coefficient tends to depend on only a few physical parameters for Hermitian case.

For the Hermitian model defined in eq. \ref{use_h}, our numerical fitting confirm that
\begin{align}
	&D^{'} \propto \frac{1}{\sqrt{2}\alpha_2} \ln(|\mu_2-\mu_2^{*}|), \quad \text{for 1D} \nonumber \\
    &D^{''} \propto \frac{2}{\alpha^2} \text{ln}(|\mu_2-\mu_2^{*}|), \quad \text{for 2D}.
\end{align}
For non-Hermitian systems, the divergence behavior is as follows:
\begin{align}
	&D^{''} \propto \frac{C_1}{\mu_2-\mu_2^{*}}+\frac{C_2}{\sqrt{|\mu_2-\mu_2^{*}|}},\quad \text{for 1D} \\
	&D^{''} \propto C_1 ln(|\mu_2-\mu_2^{*}|)+ \frac{C_2}{\sqrt{|\mu_2-\mu_2^{*}|}}, \quad \text{for 2D}. \nonumber
\end{align}

{\it Open boundary for non-Hermitian systems.-} The bulk-boundary correspondence, which was originally developed for Hermitian systems. However, for non-Hermitian systems, the open-boundary spectrum significantly differs from the periodic boundary case. Specifically, the momentum-space Hamiltonian $H(k)$ may fail to fully determine the zero modes. In general, the zero modes or phase boundaries could be observed in the numerical spectra of the real-space Hamiltonian with open boundaries. In this context, we proposed a universally applicable approach that directly employs manifold distance to determined phase boundaries in momentum space. As a specific illustrative example, we consider the 1D PT-symmetric non-Hermitian Su-Schrieffer-Heeger (SSH) model \cite{yao2018edge}.

\begin{align}
	H_k=\left(\begin{array}{cc}
		0 & \beta(k)\\
		\beta^{*}(k) & 0 \\
	\end{array}\right),
	\label{pt-nh}
\end{align}
with 
\begin{align}
	& \beta(k) = \frac{\gamma}{2} + t + (t^{'} +t_3) \cos(k) - i (t^{'}-t_3)  \sin(k), \\
	& \beta^{*}(k)= -\frac{\gamma}{2} + t + (t^{'} +t_3) \cos(k) + i (t^{'}-t_3) \sin(k) \nonumber
	\label{pt-H}
\end{align}

We show a shortcut, which is applicable only to the $t_3=0$ case. For PBCs, the phase critical lines are
\begin{equation}
t=t^{\prime} \pm (\frac{\gamma}{2}) ; \quad t=-t^{\prime} \pm(\frac{\gamma}{2}),
\label{PBCt}
\end{equation}
and the devergence behavior as follow
\begin{align}
	D^{''} \propto \frac{a}{\sqrt{|t_2-t_2^{*}|}}+b \ln(|t_2-t_2^{*}|)
\end{align}

In non-Hermitian systems, the open-boundary spectra quite different from periodic-boundary ones, which seems to indicate a complete breakdown of bulk-boundary correspondence, and its transition points
\begin{align}
	t= \pm \sqrt{ \pm(t^{\prime})^2+(\frac{\gamma}{2})^2}; \quad t= \pm \sqrt{-(t^{\prime})^2+(\frac{\gamma}{2})^2},
	\label{OBCt}
\end{align}
Nevertheless, the derivative of maifold distance $D$ can also manifest the phase transition, whether the system has PBC or OBC, as show in fig.\ref{fig3}. However, for OBC, some modifications are necessary for manifold distance.

i) The integration region for manifold distance should be the generalized Brillouin zone; more precisely, it should include the "singularities" of GBZ;

ii) Correspondingly, it is necessary to extended the momentum $k$ to its complex form, i.e.,
\begin{align}
	k \to k - i \ln r
\end{align}
which is equivalent to replacing the Bloch phase factor $e^{ik}$ by $\beta = r e^{ik}$ in OBC.

Similarly, for systems with next-nearest neighbors, we could consider the phase boundary, as shown in Fig. \ref{fig3}. While an analytical expression for the phase boundary is not available, it could be numerically obtained through the singularities of the manifold distance.

In fact, for non-lattice systems, it is necessary to truncate the integration domain, as long as these "singular points" are contained within it. In this way, its derivatives also exhibit divergence properties near phase boundary, such as p-wave superconductor model shown in fig.\ref{fig4}. Even for topological order systems with long-range entangled states, our manifold distance approach remains effective, as demonstrated in Fig. \ref{fig4}.

To conclude, we have defined the manifold distance over two manifolds and shown that its higher-order derivatives can exhibit scaling laws at the critical points when crossing the topological phase boundary. We have identified some divergence behaviors in one- and two-dimensional models and have demonstrated that this approach can be extended to non-Hermitian systems with open boundary conditions (OBCs).

For future research, we plan to extend this concept to mixed states and apply it to open systems to investigate the effects of gain and loss on various experimental platforms. Specifically, we aim to study the manifold distance given by $D = \int_{GBZ} \operatorname{Tr}\left[\sqrt{\rho^{1 / 2} \sigma \rho^{1 / 2}}\right] d \mathbf{k}$. We expect to obtain similar results for open systems. Furthermore, we aim to generalize these definitions to broader domains, such as real space or quasi-crystal systems.

{\it Acknowledgments.-} This work is supported by ...

\bibliography{ref.bib}
\end{document}